\pdfoutput=1 

\documentclass[amsmath, amssymb, amsfonts, groupedaddress, aps, prx, twocolumn,showpacs]{revtex4-1}

\usepackage{graphicx}
\usepackage{bm}
\usepackage{SIunits}
\usepackage{dcolumn}
\usepackage{tikz}
\usetikzlibrary{%
   decorations.pathreplacing,%
   decorations.pathmorphing%
}
\usepackage{color}

\renewcommand{\vec}[1]{\mathbf{{#1}}}

\newcommand{\tvec}[1]{\mathbf{#1}}
\newcommand{\pvec}[1]{{\tvec{#1}_{\parallel}}}

\newcommand{\xhat}{\tvec{\hat{x}}}

\newcommand{\kv}{\tvec{k}}
\newcommand{\qv}{\tvec{q}}

\newcommand{\xv}{\tvec{x}}
\newcommand{\xp}{\pvec{x}}
\newcommand{\kp}{\pvec{k}}
\newcommand{\qp}{\pvec{q}}

\newcommand{\Qp}{\pvec{Q}}
\newcommand{\pp}{\pvec{p}}

\newcommand{\pphat}{{\pvec{\hat{p}}}}
\newcommand{\qphat}{{\pvec{\hat{q}}}}

\newcommand{\kphat}{{\pvec{\hat{k}}}}
\newcommand{\Qm}{\mathcal{Q}}

\newcommand{\ud}{\mathrm{d}}
\newcommand{\ui}{\mathrm{i}}

\newcommand{\ve}[1]{\bm{\mathcal{E}}^\mathrm{#1}}
\newcommand{\cve}[2]{{\mathcal{E}^\mathrm{#1}_{\mathrm{#2}}}}

\newcommand{\vE}[1]{{\mathbf{E}^{\mathrm{#1}}}}

\newcommand{\vR}{\mathbf{R}}
\newcommand{\vM}{\mathbf{M}}

\newcommand{\alphas}[1]{{\alpha_{\mathrm{#1}}}}

\newcommand{\aI}{\alphas{1}}
\newcommand{\aII}{\alphas{2}}
\newcommand{\epsilons}[1]{{\varepsilon_{\mathrm{#1}}}}

\newcommand{\eI}{\epsilons{1}}
\newcommand{\eII}{\epsilons{2}}
\newcommand{\p}{\mathrm{p}}
\newcommand{\s}{\mathrm{s}}

\newcommand{\ie}{{i.e.}}

\begin{document}


\title{Numerical Simulations of Scattering of Light from Two-Dimensional Surfaces Using the Reduced Rayleigh Equation}

\author{T. Nordam}
\email[]{tor.nordam@ntnu.no}
\author{P. A. Letnes}
\email[]{paul.anton.letnes@gmail.com}
\author{I. Simonsen}
\email[]{Ingve.Simonsen@ntnu.no}
\affiliation{Department of Physics, The Norwegian University of Science and Technology (NTNU), NO-7491 Trondheim, Norway}


\begin{abstract}
  A formalism is introduced for the non-perturbative, purely numerical, solution of the reduced Rayleigh equation for the scattering of   light from two-dimensional penetrable rough surfaces. As an example,   we apply this formalism to study the scattering of p- or s-polarized light from   two-dimensional dielectric or metallic randomly rough surfaces by   calculating the full angular distribution of the co- and   cross-polarized intensity of the scattered light. In particular, we   present calculations of the mean differential reflection coefficient   for glass and silver surfaces characterized by (isotropic or   anisotropic) Gaussian and cylindrical power spectra.  The proposed   method is found, within the validity of the Rayleigh hypothesis, to   give reliable results. For a non-absorbing metal surface the   conservation of energy was explicitly checked, and found to be   satisfied to within 0.03\%, or better, for the parameters assumed.   This testifies to the accuracy of the approach and a satisfactory   discretization.
\end{abstract}

\maketitle


\section{Introduction} 
\label{sec:Introduction}

Wave scattering from rough surfaces is an old discipline which keeps attracting a great deal of attention from the scientific and technological community. Several important technologies in our society rely on such knowledge, with radar being a prime example. In the past, the interaction of light with rough surfaces was often considered an extra complication that had to be taken into account in order to properly interpret or invert scattering data. However, with the advent of nanotechnology, rough structures can be used to design novel materials with tailored optical properties. Examples include: metamaterials~\cite{MetaMaterials,AgrGar06}, photonic crystals~\cite{PhotonicCrystals}, spoof plasmons~\cite{SpoofPlasmons}, optical cloaking~\cite{Cloaking-1,Cloaking-2,Cloaking-3}, and designer surfaces~\cite{DesignerSurfaces,Simonsen2001-6}. These developments have made it even more important to have available efficient and accurate simulation tools to calculate both the far- and near-field behavior of the scattered and transmitted fields for any frequency of the incident radiation, including potential resonance frequencies of the structure.

Lord Rayleigh was the first to perform systematic studies of wave scattering from rough surfaces when, in the late 1800s, he studied the intensity distribution of a wave scattered from a sinusoidal surface~\cite{Rayleigh1907,Book:Rayleigh}. More than three decades later, Mandel'shtam studied light scattering from \emph{randomly rough} surfaces~\cite{Madelstam-1913} thereby initiating the field of wave scattering from surface disordered systems. Since the initial publication of these seminal works, numerous studies on wave scattering from randomly rough surfaces have appeared in the literature~\cite{Book:Bass-1979,Book:Ogilvy-1991,Book:Voronovich-1999,Book:Nieto-Vesperinas-2006,Book:Maradudin-2007,Zayats-2005,Simonsen-2010}, and several new multiple scattering phenomena have been predicted and confirmed experimentally. These phenomena include the enhanced backscattering and enhanced transmission phenomena, the satellite peak phenomenon, and coherent effects in the intensity-intensity correlation functions~\cite{Backscattering-1,Backscattering-2,Backscattering-3,EnhancedTransmission,Satellitepeaks,Simonsen-2010}. 

These studies, and the methods they use, can be categorized as either perturbative or purely numerical (and non-pertubative). While the former group of methods is mainly limited to weakly rough surfaces, and therefore have limited applicability, the latter group of methods can be applied to a wider class of surface roughnesses. Rigorous numerical methods can in principle be used to study the wave scattering from surfaces of any degree of surface roughness. Such simulations are routinely performed for systems where the interface has a one-dimensional roughness, i.e., where the surface structure is constant along one of the two directions of the mean plane~\cite{Maradudin1990255,Simonsen-2010}. However, for the practically more relevant situation of a two-dimensional rough surface, the purely numerical and rigorous methods are presently less used due to their computationally intensive nature. The reason for this complexity is the fact that for a randomly rough surface there is no symmetry or periodicity in the surface structure that can be used to effectively reduce the simulation domain. For a periodic surface, it is sufficient to simulate a single unit cell, while for a random surface the unit cell is in principle infinite.

A wide range of simulation methods are currently available for simulating the interaction of light with matter, including the finite-difference time-domain~(FDTD) method~\cite{FDTD}, the finite-element method~(FEM)~\cite{Book:FEM-1,Book:FEM-2}, the related surface integral equation techniques also known as the boundary element method~(BEM) or the method of moments~(MoM)~\cite{Book:BEM-1,Book:BEM-2,Book:MoM,Book:SpectralMethods,   Simonsen2010-04}, the reduced Rayleigh equation~(RRE) technique~\cite{brown1984381, Madrazo:1997uq,   Simonsen_OptComm,Simonsen2009-5,Mcgurn:1996fk,PhysRevB.63.245411,Soubret:01,Zayats-2005}, and spectral methods~\cite{Book:SpectralMethods}.

The FDTD and FEM methods discretize the whole volume of the simulation domain. Due to the complex and irregular shape of a (randomly) rough surface, it is often more convenient, and may give more accurate results (for the same level of numerical complexity)~\cite{Kern:09}, to base numerical simulations on methods where only the surface itself needs to be discretized. This is the case, for example, for the surface integral technique and the reduced Rayleigh equation methods.

The reduced Rayleigh equation is an integral equation where the unknown is either the scattering amplitude or the transmission amplitude. In the former (latter) case, one talks about the reduced Rayleigh equation for reflection (transmission). For reflection this equation was originally derived by Brown~\emph{et   al.}~\cite{brown1984381}, and subsequently by Soubret \emph{et al.} \cite{PhysRevB.63.245411,Soubret:01}. Later it has also been derived for transmission~\cite{RRE_Transmission} and film geometries~\cite{PhysRevB.63.245411,Lekova_RRE,OUR_Satellite_Paper}.

In the past, the surface integral technique has been used to study light scattering from two-dimensional randomly rough, perfectly conducting or penetrable surfaces~\cite{Simonsen2009-1,Simonsen2009-9,Simonsen2010-04}. However, to date, a direct numerical and non-perturbative solution of the two-dimensional reduced Rayleigh equation has not appeared in the literature, even if its one-dimensional analog has been solved numerically and has been used to study the scattering from, and transmission through, one-dimensional rough surfaces~\cite{Madrazo:1997uq,Simonsen_OptComm,Simonsen2009-5}. The lesson learned from the one-dimensional scattering studies reported in Refs.~\cite{Madrazo:1997uq,Simonsen_OptComm,Simonsen2009-5} is that simulations based on a direct numerical solution of the reduced Rayleigh equation may give accurate non-perturbative results for systems where alternative methods struggle to give the same level of accuracy. Moreover, the reduced Rayleigh method also requires less memory for the same surface dimensions when compared to, e.g., the rigorous surface integral technique.
 
The main aim of this paper is to present a numerical method and formalism for the solution of the two-dimensional reduced Rayleigh equation for reflection. While we exclusively consider reflection, the formalism for transmission will be almost identical, and the resulting equation will have a similar form as for reflection. Additionally, the equation for transmission or reflection for a film geometry, i.e., for a film of finite thickness on top of a substrate, where only one interface is rough, will also have a similar form. The method presented will be illustrated by applying it to the study of the scattering of p- or s-polarized light from two-dimensional metallic or dielectric media separated from vacuum by an isotropic or anisotropic randomly rough surface.

This paper is organized as follows: First, in Sec.~\ref{sec:geometry} we present the scattering geometry to be considered. We will then present some relevant scattering theory, including the reduced Rayleigh equation for the geometry under study (Sec.~\ref{sec:scatteringtheory}), followed by a detailed description of how the equation can be  solved numerically (Sec.~\ref{sec:solving}). Next, we will present some simulation results obtained by the introduced method~(Sec.~\ref{sec:results}). We then discuss some of the computational challenges of this method~(Sec.~\ref{sec:discussion}), and, finally, in Sec.~\ref{sec:Conclusion} we draw some conclusions.

\section{Scattering Geometry} 
\label{sec:geometry}

\begin{figure}
  \includegraphics{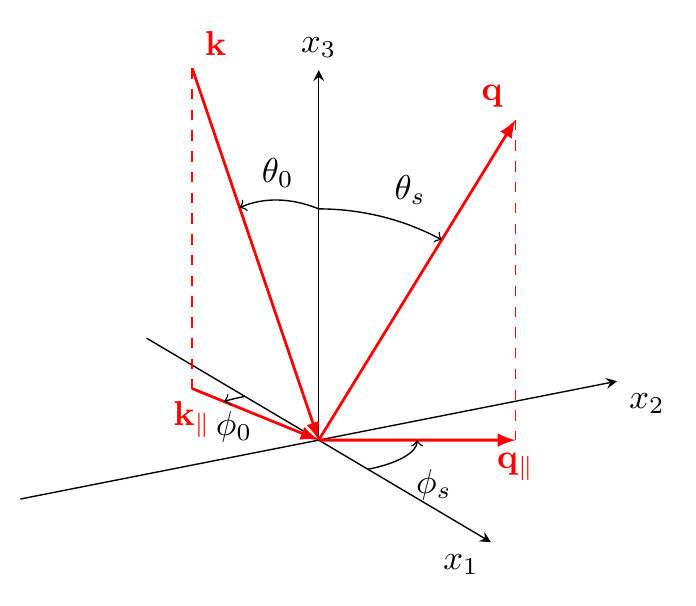}
  \caption{(Color online) A sketch of the scattering geometry assumed in this work. The figure also shows the coordinate system used, angles of incidence $(\theta_0,\phi_0)$ and scattering $(\theta_s,\phi_s)$, and the corresponding lateral wavevectors $\pvec{k}$ and $\pvec{q}$, respectively.}
  \label{fig:geometry}
\end{figure}

We consider a system where a rough surface separates two regions. Region~1 is assumed to be vacuum ($\eI = 1$), and region~2 is filled with a metal or dielectric characterized by a complex dielectric function $\eII(\omega)$, where the angular frequency is $\omega=2\pi c/\lambda$, with $\lambda$ being the wavelength of the incident light in vacuum and $c$ the speed of light in vacuum. The height of the surface measured in the positive $x_3$ direction from the $x_1 x_2$-plane is given by the single-valued function $x_3=\zeta(\xp)$, where $\xp=(x_1, x_2, 0)$, which is assumed to be at least once differentiable with respect to $x_1$ and $x_2$. Angles of incidence $(\theta_0,\phi_0)$ and scattering $(\theta_s,\phi_s)$ are defined positive according to the convention given in Fig.~\ref{fig:geometry}.

In principle, the theory to be presented in Sec.~\ref{sec:scatteringtheory} can be used to calculate the scattering of light from any surface, provided it is not too rough. However, in this paper, we will consider randomly rough surfaces where $\zeta(\xp)$ constitutes a stationary random process defined by
\begin{align}
\begin{aligned}
\label{eq:surface_definition}
    \left\langle\zeta(\xp)\right\rangle
&=
    0,
\\
    \left\langle
        \zeta(\xp)
        \zeta(\xp')
    \right\rangle
&=
    \delta^2
    W(\xp-\xp'),
\end{aligned}
\end{align}
where the angle brackets denote an average over an ensamble of surface realizations.  In writing Eqs.~(\ref{eq:surface_definition}) we have defined the root-mean-square height of the surface, $\delta=\left<\zeta^2(\pvec{x})\right>^{1/2}$, and $W(\xp-\xp')$ denotes the height-height auto-correlation function of the surface, normalized so that $W(\vec{0})=1$~\cite{Simonsen-2010}.  According to the Wiener-Khinchin theorem~\cite{Book:VanKampen-2007}, the power spectrum of the surface profile function is given by
\begin{align}
\label{eq:powerspectrum}
 g(\pvec{k} ) 
     &= 
      \int\! \mathrm{d}^2x_\parallel\; 
      W(\pvec{x} ) \exp\left(-\mathrm{i}\pvec{k}\cdot\pvec{x} \right).
\end{align}
The power spectra that will be considered in this work are of either the Gaussian form~\cite{Simonsen2010-04}
\begin{align}
\label{eq:gaussian}
g(\kp ) &= \sqrt{\pi}a_1a_2
            \exp\left( -\frac{k_1^2 a_1^2}{4}
                       -\frac{k_2^2 a_2^2}{4}
               \right),
\end{align}
where $a_i$ ($i=1,2$) denotes the lateral correlation length for
direction $i$, or the cylindrical form~\cite{Mcgurn:1996fk}
\begin{align}
\label{eq:cylindrical}
    \begin{aligned}
    g(k_\parallel)  =& \frac{4\pi}{k_+^2 - k_-^2}\left[\theta(k_\parallel-k_-)\theta(k_+ - k_\parallel)\right],
\end{aligned}
\end{align}
where $k_\parallel=|\kp|$, $\theta$ denotes the Heaviside unit step function, and $k_\pm$ are wavenumber cutoff parameters, with $k_-<k_+$. The cylindrical form in Eq.~(\ref{eq:cylindrical}) is a two-dimensional generalization of the power spectrum used in the experiments where West and O'Donnell confirmed the existence of the enhanced backscattering phenomenon for weakly rough surfaces~\cite{Backscattering-2}.


\section{Scattering Theory} 
\label{sec:scatteringtheory}

We consider a linearly p- or s-polarized plane wave
which is incident on the surface from region~1, with the electric
field given by $\vE{inc}(\xv;t)=\vE{inc}(\xv|\omega) \exp(-\ui\omega t)$
where
\begin{subequations}
  \label{eq:E_inc_total}
\begin{align}
\label{eq:E_inc}
    \vE{inc}(\xv| \omega)
    ={}&
    \ve{inc}(\kp)\exp\left[\ui\kp\cdot\xp-\ui\aI(k_\parallel)x_3\right],
\end{align}
with
\begin{align}
  \label{eq:E_inc_amplitudes}
\begin{aligned}
    \ve{inc}(\kp)
    ={}&
    -\frac{c}{\omega}
    \left[
        \kphat\aI(k_\parallel)+\xhat_3 k_\parallel
    \right]
    \cve{inc}{p}(\kp)
\\
&+
    \left(\xhat_3 \times \kphat\right)\cve{inc}{s}(\kp),
\end{aligned}
\end{align}
and
\begin{align}
  \label{eq:alpha_1}
  \aI(k_\parallel) 
&=
\sqrt{\frac{\omega^2}{c^2}-k_\parallel^2}
    ,\;\; \mathrm{Re}\,\aI \geq 0,\;\mathrm{Im}\,\aI \geq 0.
\end{align}
\end{subequations}
Here, and in the rest of the paper, a caret over a vector indicates a unit vector.  The expressions in front of the amplitudes $\cve{inc}{\alpha}(\kp)$ ($\alpha=\mathrm{p,s}$) in Eq.~(\ref{eq:E_inc_amplitudes}) correspond to unit polarization vectors for incident light of linear polarization $\alpha$. Moreover, $\kp = (k_1, k_2, 0)$ denotes the lateral component of the wave vector $\vec{k}=\pvec{k}-\alpha(k_\parallel)\vec{\hat{x}}_3$. When the lateral wavenumber satisfies $k_\parallel \leq \omega/c$, as will be assumed here, $\pvec{k}$ is related to the angles of incidence according to
\begin{align}
  \label{eq:k-parallel}
  \pvec{k} &= 
  \frac{\omega}{c} \sin\theta_0 \left(\cos\phi_0, \sin\phi_0, 0 \right),
\end{align}
where $c$ denotes the speed of light in vacuum and $\theta_0$ and $\phi_0$ are the polar and azimuthal angles of incidence, respectively~(Fig.~\ref{fig:geometry}).  When writing the field of incidence, $\vE{inc}(\xv;t)$, a time harmonic dependence of the form $\exp(-\ui\omega t)$ was assumed. A similar time dependence will be assumed for all field expressions, but not indicted explicitly.

Above the surface roughness region, i.e., for $x_3 > \max \zeta(\xp)$, the scattered field can be written as a superposition of {\em upwards} propagating reflected plane waves:
\begin{subequations}
  \label{eq:E_sc_total}
\begin{align}
\begin{aligned}
\label{eq:E_sc}
    \vE{sc}(\xv| \omega)
    ={}& \int \frac{\mathrm{d}^2q_\parallel}{(2\pi)^2}
    \ve{sc}(\qp)
\\
    &\times\exp\left[\ui\qp\cdot\xp+\ui\aI(q_\parallel)x_3\right],
\end{aligned}
\end{align}
where
\begin{align}
\begin{aligned}
    \ve{sc}(\qp)
    ={}&
   \frac{c}{\omega}
    \left[
        \qphat\aI(q_\parallel)-\xhat_3 q_\parallel
    \right]
    \cve{sc}{p}(\pvec{q})
\\
&+ 
    \left(\xhat_3 \times \qphat\right)\cve{sc}{s}(\pvec{q}).
\end{aligned}
\end{align}
\end{subequations}
The integration in Eq.~(\ref{eq:E_sc}) is over the entire plane, including the evanescent region \mbox{$q_\parallel>\omega/c$}. Therefore, both propagating and evanescent modes are included in $\vE{sc}(\xv | \omega)$.

We will assume that a linear relationship exists between the amplitudes of the incident and the scattered fields, and we write (for $\alpha=\mathrm{p,s}$)
\begin{align}
\label{eq:scattering_amplitude}
    \cve{sc}{\alpha}(\qp) 
    ={}&
    \sum_{\beta=\p,\s}
    R_{\alpha\beta}(\qp|\kp)
    \cve{inc}{\beta}(\kp).
\end{align}
Here we have introduced the so-called \emph{scattering amplitude} $R_{\alpha\beta}(\qp|\kp)$, which describes how incident $\beta$-polarized light characterized by a lateral wave vector $\pvec{k}$ is converted by the surface roughness into scattered light of polarization $\alpha$ and lateral wave vector $\pvec{q}$. When $q_\parallel \leq \omega/c$, the wave vector $\pvec{q}$ is related to the angles of scattering $(\theta_s,\phi_s)$ by
\begin{align} 
  \label{eq:q-parallel}
  \pvec{q} &= \frac{\omega}{c} \sin\theta_s 
        \left(\cos\phi_s, \sin\phi_s, 0 \right).
\end{align} 

Below the surface region, i.e., for $x_3< \min\,\zeta(\pvec{x})$, the transmitted electric field can be written as 
\begin{subequations}
  \label{eq:E_tr_total}
\begin{align}
  \label{eq:E_tr}
  \begin{aligned}    
      \vE{tr}(\pvec{x}|\omega) 
      ={}&
      \int \frac{\ud^2 p_\parallel}{(2\pi)^2}
      \ve{tr}(\pvec{p})
\\
      &\times\exp \left[ \ui\pvec{p}\cdot \xp - \ui\aII(p_\parallel) x_3 \right]
  \end{aligned}
\end{align}
with
\begin{align}
  \label{eq:E_tr_amplitudes}
\begin{aligned}
    \ve{tr}(\pvec{p})
    ={}&
    -\frac{1}{\sqrt{\varepsilon_2(\omega)}}
    \frac{c}{\omega}
    \left[
        \vec{\hat{p}}_\parallel \aII(p_\parallel)+\xhat_3 p_\parallel
    \right]
    \cve{tr}{p}(\pvec{p})
\\
&+
    \left(\xhat_3 \times \vec{\hat{p}}_\parallel \right)\cve{tr}{s}(\pvec{p}).
\end{aligned}
\end{align}
\end{subequations}
In writing Eqs.~(\ref{eq:E_tr_total}) we have introduced wave vectors of the transmitted field $\vec{p}=\pvec{p}-\aII(p_\parallel)\vec{\hat{x}}_3$, where 
\begin{align}
  \label{eq:alpha_2}
  \begin{aligned}
      \aII(p_\parallel) 
  &{}=
  \sqrt{\varepsilon_2(\omega)\frac{\omega^2}{c^2}-p_\parallel^2},\\
  &\mathrm{Re}\,\aII \geq 0,\;\mathrm{Im}\,\aII \geq 0.
\end{aligned}
\end{align}
In complete analogy to what was done for reflection, a transmission amplitude $T_{\alpha\beta}(\pvec{p}|\pvec{k})$ may be defined via the following linear relation between the amplitudes of the incident and transmitted fields ($\alpha=\mathrm{p,s}$)
\begin{align}
  \label{eq:transmission_amplitude}
  \cve{tr}{\alpha}(\pvec{p}) 
  ={}& 
  \sum_{\beta=\p,\s}
  T_{\alpha\beta}(\pvec{p}|\pvec{k})
  \cve{inc}{\beta}(\kp).
\end{align}

Since the form of the electric fields given by Eqs.~\eqref{eq:E_inc_total}, \eqref{eq:E_sc_total}, and \eqref{eq:E_tr_total} apply far away from the surface region, they are referred to as the {\em asymptotic forms} of the electric field. These equations automatically satisfy the boundary conditions at infinity.

In passing we note that once the incident field has been specified, the scattered and transmitted fields are fully specified outside the surface roughness region if the reflection [$R_{\alpha\beta}(\pvec{q}|\pvec{k})$] and transmission [$T_{\alpha\beta}(\pvec{p}|\pvec{k})$] amplitudes are known. We will now address how the reflection amplitude can be calculated.

\subsection{The Rayleigh Hypothesis}

Above the surface, i.e., in the region $x_3 > \max \zeta(\xp)$, the total electric field is equal to the sum of the incident and the scattered field, $\vE{inc}(\xv |\omega) + \vE{sc}(\xv | \omega)$. Below the surface, in the region $x_3 < \min\zeta(\xp)$, it equals the transmitted field, $\vE{tr}(\xv | \omega)$. In the surface roughness region, $\min \zeta(\xp) \leq x_3 \leq \max \zeta(\xp)$, these forms of the total field will not generally be valid. In particular, when we are above the surface but still below its maximum point, i.e., $\zeta(\xp) \leq x_3< \max\zeta(\xp)$, the expression for the scattered field will also have terms containing $\exp\left[\ui\qp\cdot\xp-\ui\aI(q_\parallel)x_3\right]$. Similarly, the transmitted field in the surface region has to contain an additional term similar to Eq.~\eqref{eq:E_tr} but with the exponential function replaced by $\exp\left[\ui\qp\cdot\xp+\ui\aII(q_\parallel)x_3\right]$ (and a different amplitude).

If the surface roughness is sufficiently weak, however, the asymptotic form of the fields, Eqs.~\eqref{eq:E_inc_total}, \eqref{eq:E_sc_total}, and \eqref{eq:E_tr_total}, can be assumed to be a good approximation to the total electric field in the surface roughness region. This assumption is known as the \emph{Rayleigh   hypothesis}~\cite{Book:Rayleigh,Rayleigh1907,Book:Maradudin-2007}, in honor of Lord Rayleigh, who first used it in his seminal studies of wave scattering from sinusoidal surfaces~\cite{Book:Rayleigh,Rayleigh1907}. For a (one-dimensional) sinusoidal surface, $x_3=\zeta_0\sin(\Lambda x_1)$, the criterion for the validity of the Rayleigh hypothesis, and thus equations that can be derived from it (like the reduced Rayleigh equation to be introduced below), is known to be $\zeta_0\Lambda<0.448$, independent of the wavelength of the incident light~\cite{Millar-1969,Millar-1971}. For a randomly rough surface, however, the absolute limit of validity of this hypothesis is not generally known, though some numerical studies have been devoted to finding the region of validity for random surfaces~\cite{Tishchenko-2009}. Even if no absolute criterion for the validity of the Rayleigh hypothesis for randomly rough surfaces is known, it remains true that it is a small-slope hypothesis. In particular, if the randomly rough surface is characterized by an rms height $\delta$, and a correlation length $a$ (see Sec.~\ref{sec:geometry} and Ref.~\cite{Simonsen-2010} for details), there seems to be a consensus in the literature on the Rayleigh hypothesis being valid if $\delta/a\ll 1$~\cite{Tishchenko-2009,Book:Maradudin-2007}. We stress that the validity of the Rayleigh hypothesis does not require the amplitude of the surface roughness to be small, only its slope.

\subsection{The Reduced Rayleigh Equations}

Under the assumption that the Rayleigh hypothesis is valid, the total electric field in the surface region, $\min \zeta(\xp) < x_3 < \max \zeta(\xp)$, can be written in the form given by Eqs.~\eqref{eq:E_inc_total}, \eqref{eq:E_sc_total} and \eqref{eq:E_tr_total} [with Eqs.~\eqref{eq:scattering_amplitude} and \eqref{eq:transmission_amplitude}]. Hence, these asymptotic fields can be used to satisfy the usual boundary conditions on the electromagnetic field at the rough surface $x_3=\zeta(\pvec{x})$~\cite{Book:Jackson-2007,Book:Stratton-2007}. In this way, one obtains the so-called Rayleigh equations, a set of coupled inhomogeneous integral equations, which the reflection and transmission amplitudes should satisfy.

In the mid-1980s, it was demonstrated by Brown \emph{et   al.}~\cite{brown1984381} that either the reflection or transmission amplitude could be eliminated from the Rayleigh equations, resulting in an integral equation for the remaining amplitude only. Since this latter integral equation contains only the field above (below) the rough surface, it has been termed the \emph{reduced Rayleigh equation} for reflection (transmission). Subsequently, reduced Rayleigh equations for two-dimensional film geometries, i.e., a film of finite thickness on top of an infinitely thick substrate, where only one interface is rough, was derived by Soubret \emph{et   al.}~\cite{PhysRevB.63.245411,Soubret:01} and Leskova~\cite{Lekova_RRE,OUR_Satellite_Paper}. Moreover, reduced Rayleigh equations for reflection from clean, perfectly conducting, two-dimensional randomly rough surfaces~\cite{RRE_PEC} and reduced Rayleigh equations for transmission through clean, penetrable two-dimensional surfaces~\cite{RRE_Transmission} have been derived.

For the purposes of the present study, we limit ourselves to a scattering system consisting of a clean, penetrable, two-dimensional rough surface $x_3=\zeta(\pvec{x})$ (Sec.~\ref{sec:geometry}).  If the scattering amplitudes are organized as the $2\times2$ matrix
\begin{align}
    \vR(\qp|\kp)
    &=
    \left(\begin{array}{cc}
        R_{pp}(\qp|\kp) & R_{ps}(\qp|\kp) \\
        R_{sp}(\qp|\kp) & R_{ss}(\qp|\kp)
    \end{array}\right),
\end{align}
the reduced Rayleigh equation (for reflection) for this geometry can be written in the form~\cite{Mcgurn:1996fk,PhysRevB.63.245411,Soubret:01}
\begin{subequations}
  \label{eq:RRE-total}
\begin{widetext}
\begin{align}
\label{eq:RRE}
	\int\frac{\ud^2q_\parallel}{(2\pi)^2}
	\frac
        {I\left(\aII(\pp)-\aI(\qp)|\pp-\qp\right)}
		{\aII(\pp)-\aI(\qp)}
	\vM^{+}(\pp|\qp)
	\vR(\qp|\kp)
=
	-\frac
        {I\left(\aII(\pp)+\aI(\kp)|\pp-\kp\right)}
		{\aII(\pp)+\aI(\kp)}
    \vM^{-}(\pp|\kp),
\end{align}
where
\begin{align}
\label{eq:I_integral}
   I(\gamma|\Qp)
   =&
   \int\ud^2x_\parallel
   \exp\left[-\ui\gamma\zeta(\xp)\right]
   \exp\left(-\ui\Qp\cdot\xp\right), 
\end{align}
and
\begin{align}
   \vM^\pm(\pp|\qp)
   =
	\left(
		\begin{array}{cc}
            p_\parallel q_\parallel \pm \aI(\qp)\aII(\pp)\pphat \cdot \qphat &
            -\frac{\omega}{c}\aII(\pp)\left[\pphat \times \qphat\right]_3 \\
            \pm\frac{\omega}{c}\aI(\qp)\left[\pphat\times\qphat\right]_3 &
				\frac{\omega^2}{c^2}\pphat\cdot\qphat
		\end{array}
	\right),
\end{align}
\end{widetext}
\end{subequations}
where the integrals in Eqs.~(\ref{eq:RRE}) and (\ref{eq:I_integral}) are over the entire $\qp$-plane and $\xp$-plane, respectively. Reduced Rayleigh equations for transmission, or film geometries with only one rough interface, will have a similar structure to Eq.~\eqref{eq:RRE-total}~\cite{PhysRevB.63.245411,Soubret:01}, and can be solved in a completely analogous fashion.

It should be mentioned that the reduced Rayleigh equation can serve as a starting point for most, if not all, perturbation theoretical approaches to the study of scattering from rough surfaces~\cite{Simonsen-2010}. For example, McGurn and Maradudin studied the scattering of light from two-dimensional rough surfaces based on the reduced Rayleigh equation, going to fourth order in the expansion in the surface profile function, and demonstrating the presence of enhanced backscattering~\cite{Mcgurn:1996fk}.  

\subsection{Mean Differential Reflection Coefficient} 
\label{sub:mdrc}
The solution of the reduced Rayleigh equation determines the scattering amplitudes $R_{\alpha\beta}(\qp | \kp)$. While this quantity completely specifies the total field in the region above the surface, it is not directly measurable in experiments. A more useful quantity is the mean differential reflection coefficient (DRC), which is defined as the time-averaged fraction of the incident power scattered into the solid angle $\ud \Omega_s$ about the scattering direction $\qv$. The mean DRC is defined as~\cite{Mcgurn:1996fk}
\begin{align}
    \label{eq:drc}
    \left\langle
        \frac{\partial R_{\alpha\beta}}{\partial \Omega_s}
    \right\rangle
=
    \frac{1}{L^2}
    \frac{\omega^2}{4 \pi^2 c^2}
    \frac{\cos^2 \theta_s}{\cos \theta_0}
    \left<
        \left| R_{\alpha\beta}(\qp | \kp) \right|^2
    \right>,
\end{align}
where $L^2$ is the area of the plane $x_3=0$ covered by the rough surface. In this work, we are mainly interested in diffuse (incoherent) scattering. Since we consider weakly rough surfaces, the specular (coherent) scattering will dominate, and it will be convenient to separate the mean DRC into its coherent and incoherent parts. By coherent scattering, we mean the part of the scattered light which does not cancel when the ensemble average of $R_{\alpha\beta}$ is taken, i.e., the part where the scattered field is in phase between surface realizations. Conversely, the incoherent part is the part which cancels in the ensemble average. The component of the mean DRC from incoherent scattering is~\cite{Mcgurn:1996fk}
\begin{align}
\begin{aligned}
    \label{eq:drc_incoh}
   &\left<
        \frac{\partial R_{\alpha\beta}}{\partial \Omega_s}
        \right>_{\text{incoh}}
      =
    \frac{1}{L^2}
    \frac{\omega^2}{4 \pi^2 c^2}
    \frac{\cos^2 \theta_s}{\cos \theta_0}
    \\
   &\qquad \times
   \left[\left<
        \left| R_{\alpha\beta}(\qp | \kp) \right|^2
    \right>
    -\left|\left<
        R_{\alpha\beta}(\qp | \kp)
    \right> \right|^2
\right].
\end{aligned}
\end{align}
The contribution to the mean DRC from the coherently scattered light is given by the difference between Eqs.~\eqref{eq:drc} and \eqref{eq:drc_incoh}.

\subsection{Conservation of Energy} 
\label{sub:energy}
As a way to check the accuracy of our results, it is useful to investigate energy conservation. If we consider a metallic substrate with no absorption, the reflected power should be equal to the incident power. The fraction of the incident light of polarization $\beta$ which is scattered into polarization $\alpha$ is given by the integral of the corresponding mean DRC over the upper hemisphere:
\begin{align}
    \mathcal{U}_{\alpha\beta}
=&
    \int \ud \Omega_s\;
    \left\langle
        \frac{\partial R_{\alpha\beta}}{\partial \Omega_s}
    \right\rangle.
\end{align}
For a non-absorbing metal, if we send in light of polarization $\beta$, we should have
\begin{align}
    \sum_{\alpha} \mathcal{U}_{\alpha\beta} = 1,
\end{align}
if energy is conserved. While the conservation of energy is useful as a relatively simple test, it is important to note that it is a necessary, but not sufficient, condition for correct results.

\section{Numerical Solution of the Reduced Rayleigh Equation} 
\label{sec:solving}
The starting point for the numerical solution of the reduced Rayleigh equation is a discretely sampled surface, from which we wish to calculate the reflection. We will limit our discussion to quadratic surfaces of size $L \times L$, sampled on a quadratic grid of $N_x \times N_x$ points with a grid constant
\begin{align}
\label{eq:deltax}
    \Delta x
=
    \frac{L}{N_x}.
\end{align}

In this paper, we will present results for numerically generated random surfaces. These are generated by what is known as the Fourier filtering method. Briefly, it consists of generating uncorrelated random numbers with a Gaussian distribution, transforming them to Fourier space, filtering them with the square root of the surface power spectrum $g(\kp)$, and transforming them back to real space. The interested reader is referred to, e.g., Refs.~\cite{Simonsen2010-04,Maradudin1990255}.

The next step towards the numerical solution of the reduced Rayleigh equation is the evaluation of the integrals $I(\gamma|\pvec{Q})$ defined in Eq.~\eqref{eq:I_integral}. These integrals are so-called Fourier integrals and care should be taken when evaluating them due to the oscillating integrands~\cite{Book:NR-1992}. Using direct numerical integration routines for their evaluation will typically result in inaccurate results. Instead, a (fast) Fourier transform technique with end point corrections may be adapted for their evaluation, and the details of the method is outlined in Ref.~\cite{Book:NR-1992}. However, these calculations are time consuming, since $I(\gamma|\pvec{Q})$ must be evaluated for all values of the arguments $\gamma=\aI(\pp)-\aII(\qp)$ and $\gamma=\aI(\pp)-\aII(\kp)$~\footnote{For the calculations used to generate the results presented in this paper, this would amount to evaluating $I(\gamma|\Qp)$ on the order of $10^{10}$ times.}.

Instead, a computationally more efficient way of evaluating $I(\gamma|\Qp)$ is to assume that the exponential function $\exp\left[-\ui\gamma\zeta(\xp)\right]$, present in the definition of $I(\gamma|\pvec{Q})$, can be expanded in powers of the surface profile function, and then evaluating the resulting expression term-by-term by Fourier transform. This gives
\begin{subequations}
  \label{eq:I_fourier_expansion}
\begin{align}
\label{eq:I_fourier_expansion_A}
   I(\gamma|\Qp)
   =&	\sum_{n=0}^{\infty}
	\frac{(-\ui\gamma)^{n}}{n!}
        \hat{\zeta}^{(n)}(\pvec{Q}),
\end{align}
where $\hat{\zeta}^{(n)}(\pvec{Q})$ denotes the Fourier transform of the $n$th power of the profile function, \ie,
\begin{align}
  \label{eq:I_fourier_expansion_B}
  \hat{\zeta}^{(n)}(\pvec{Q})
  =&
  \int\ud^2 x_\parallel
  \zeta^{n}(\xp)
  \exp\left(-\ui\Qp\cdot\xp\right). 
\end{align}
\end{subequations}
In practice, the sum in Eq.~\eqref{eq:I_fourier_expansion_A} will be truncated at a finite value $n=J$, and the Fourier transforms are calculated using a fast Fourier transform (FFT) algorithm.

The advantage of using Eqs.~\eqref{eq:I_fourier_expansion} for calculating $I(\gamma|\Qp)$, rather than the method of Ref.~\cite{Book:NR-1992}, is that the Fourier transform of each power of $\zeta(\pvec{x})$ can be performed once, and changing the argument $\gamma$ in $I(\gamma|\pvec{Q})$ will not require additional Fourier transforms to be evaluated. This results in a significant reduction in computational time. The same method has previously been applied successfully to the numerical solution of the one-dimensional reduced Rayleigh equation~\cite{Madrazo:1997uq,Simonsen_OptComm,Simonsen2009-5}.

It should be noted that the Taylor expansion used to arrive at Eq.~\eqref{eq:I_fourier_expansion} requires that $\left|\gamma\zeta(\pvec{x})\right|\ll 1$ to converge reasonably fast, putting additional constraints on the amplitude of the surface roughness which may be more restrictive than those introduced by the Rayleigh hypothesis. Hence, surfaces exist for which the Rayleigh hypothesis is satisfied, but the above expansion method will not converge, and the more time-consuming approach of Ref.~\cite{Book:NR-1992} will have to be applied. 


Next, we need to truncate and discretize the integral over $\qp$ in Eq.~(\ref{eq:RRE}). We discretize $\qp$ on a grid of equidistant points, with spacing $\Delta q$, such that
\begin{align}
\label{eq:qgrid}
    \qp_{ij}=\left(-\frac{\Qm}{2}+i\Delta{}q, -\frac{\Qm}{2}+j\Delta{}q, 0\right),
\end{align}
where $i, j = 0,1,2,\ldots, N_q-1$, and $\Qm=\Delta q (N_q-1)$. Here, $N_q$ denotes the number of points along each axis of the grid. Additionally, we limit the integration over $\qp$ to the region $q_\parallel \leq \Qm/2$. The choice of a circular integration domain reduces the computational cost, and will be discussed in more detail in Sec.~\ref{sec:discussion}. Converting the integral into a sum by using a two-dimensional version of the standard mid-point quadrature scheme, we get the equation:
\begin{widetext}
\begin{align}
\begin{aligned}
\label{eq:RRE2}
    \left(\frac{\Delta{}q}{2\pi} \right)^2
    \sum_{{q_\parallel}_{ij} \leq \mathcal{Q}/2}
&    \frac
    {I\left(\aII(\pp)-\aI(\qp_{ij})|\pp-\qp_{ij}\right)}
    {\aII(\pp)-\aI(\qp_{ij})}
    \vM^{+}(\pp|\qp_{ij})
    \vR(\qp_{ij}|\kp)
=\\
&-\frac
    {I\left(\aII(\pp)+\aI(\kp)|\pp-\kp\right)}
    {\aII(\pp)+\aI(\kp)}
    \vM^{-}(\pp|\kp).
\end{aligned} 
\end{align} 
\end{widetext}
Here, the sum is to be taken over all $\qp_{ij}$ such that ${q_\parallel}_{ij} \leq \Qm/2$, where ${q_\parallel}_{ij} = \left|\qp_{ij}\right|$. This sum yields a matrix equation where the unknowns are the four components of $\vR(\qp_{ij}|\kp)$. It is evident from Eq.~(\ref{eq:scattering_amplitude}) that if we consider incident light of either p or s polarization, we need only calculate two of the components of the scattering amplitude to fully specify the reflected field. Hence, we solve separately for either p-polarized incident light, i.e., $R_{pp}$ and $R_{sp}$, or s-polarized incident light, i.e., $R_{ss}$ and $R_{ps}$. In either case, we have twice as many unknowns as the number of values of $\qp_{ij}$ included in the sum in Eq.~(\ref{eq:RRE2}). Note that the left hand side of the equation system is the same for both incident polarizations, and will also remain the same for all angles of incidence, as $\kp$ only enters at the right hand side of Eq.~(\ref{eq:RRE2}).

In order to solve for all unknowns, we need to discretize $\pp$ as well, to obtain a closed set of linear equations. Using the same grid for $\pp$ as for $\qp$ will give us the necessary number of equations, as Eq.\eqref{eq:RRE2} yields two equations for each value of $\pp$. Since we integrate over a circular $\qp$ domain, with $\qp$ discretized on a quadratic grid, the exact number of values of $\qp_{ij}$ will depend on the particular values of $\Qm$ and $N_q$, but will be approximately $(\pi/4) N_q^2$.

In order to take advantage of the method for calculating $I(\gamma|\Qp)$ described by Eq.~(\ref{eq:I_fourier_expansion}), it is essential that all possible values of $\pp-\qp$ and $\pp-\kp$ [see Eq.~\eqref{eq:RRE2}] fall on the grid of wave vectors $\Qp$ resolved by the Fourier transform of the surface profile we used in that calculation. First, we note that when $\pp$ and $\qp$ are discretized on the same quadratic grid, the number of possible values for each component of $\pp - \qp$ will always be an odd number, $2N_q - 1$, where $N_q$ is the number of possible values for each component of $\pp$ and $\qp$. Thus, by choosing $N_q$ such that $2N_q-1$ equals the number of elements along each axis of the FFT of the surface profile we used to calculate the integrals in Eq.~\eqref{eq:I_fourier_expansion_B}, we ensure that the required number of points is resolved by the FFT \footnote{Note that since the FFT always resolves the zero frequency, and the FFT of a purely real signal is symmetric about the zero frequency under complex conjugation, it is always possible to calculate an odd number of elements along each axis of the FFT}. Hence, we choose
\begin{align}
\label{eq:Nq}
    N_q = \left\lfloor \frac{N_x+2}{2} \right\rfloor,
\end{align}
where $\lfloor x \rfloor$ is the floor function of $x$, which is equal to the largest integer less than or equal to $x$.

Next, we let $\Delta q$ equal the resolution of the FFT \cite{Book:NR-1992}, i.e.,
\begin{align}
\label{eq:deltaq}
    \Delta q = \frac{2\pi}{L}
\end{align}
and we let $\Qm$ be equal to the highest wavenumber resolved by the FFT \cite{Book:NR-1992},
\begin{align}
\label{eq:Q}
    \Qm = \Delta q \lfloor N_x/2 \rfloor.
\end{align}

In the end, we get the equation
\begin{widetext}
\begin{align}
\begin{aligned}
\label{eq:RRE3}
    \left(\frac{\Delta{}q}{2\pi} \right)^2
    \sum_{\left|\qp_{ij}\right| \leq \mathcal{Q}/2}
&    \frac
    {I\left(\aII(\pp_{kl})-\aI(\qp_{ij})|\pp_{kl}-\qp_{ij}\right)}
    {\aII(\pp_{kl})-\aI(\qp_{ij})}
    \vM^{+}(\pp_{kl}|\qp_{ij})
    \vR(\qp_{ij}|\kp_{mn})
=\\
&-\frac
    {I\left(\aII(\pp_{kl})+\aI(\kp_{mn})|\pp_{kl}-\kp_{mn}\right)}
    {\aII(\pp_{kl})+\aI(\kp_{mn})}
    \vM^{-}(\pp_{kl}|\kp_{mn}),
\end{aligned} 
\end{align} 
\end{widetext}
where $\qp_{ij}$, as well as $\pp_{kl}$ and $\kp_{mn}$, are defined on the grid given by Eq.~(\ref{eq:qgrid}), with $i,j,k,l,m,n=0,1,2,\ldots, N_q-1$, and where $N_q$, $\Delta q$ and $\Qm$ are given by Eqs.~(\ref{eq:Nq}), (\ref{eq:deltaq}) and (\ref{eq:Q}), respectively.

Evaluating Eq.~(\ref{eq:RRE3}) for all values of $\pp_{kl}$ satisfying ${p_\parallel}_{kl} \leq \mathcal{Q}/2$, and assuming one value of $\kp_{mn}$, such that ${k_\parallel}_{mn} < \omega/c$, and one incident polarization $\beta$, results in a \emph{closed} system of linear equations in $R_{\alpha\beta}(\qp_{ij}|\kp_{mn})$ where $\alpha=\p,\s$. Repeating the procedure for both incident polarizations allows us to obtain all four components of $\vR(\qp_{ij}|\kp_{mn})$.

With the reflection amplitudes $R_{\alpha\beta}(\pvec{q}_{ij}|\pvec{k}_{mn})$ available, the contribution to the mean differential reflection coefficient from the light that has been scattered incoherently is obtained from Eq.~(\ref{eq:drc_incoh}) after averaging over an ensemble of surface realizations.


In passing we note that to avoid loss of numerical precision by operating on numbers with widely different orders of magnitude, we have rescaled all quantities in our problem to dimensionless numbers. When considering an incoming wave of wavelength $\lambda$, angular frequency $\omega$, and wave vector $\kv$, we have chosen to rescale all lengths in our problem by multiplying with $\omega/c$, and all wavenumbers by multiplying with $c/\omega$, effectively measuring all lengths in units of $\lambda/2\pi$, and the magnitude of wave vectors in units of $\omega/c$.


\section{Results} 
\label{sec:results}

\begin{figure}
    \includegraphics[width=\columnwidth]{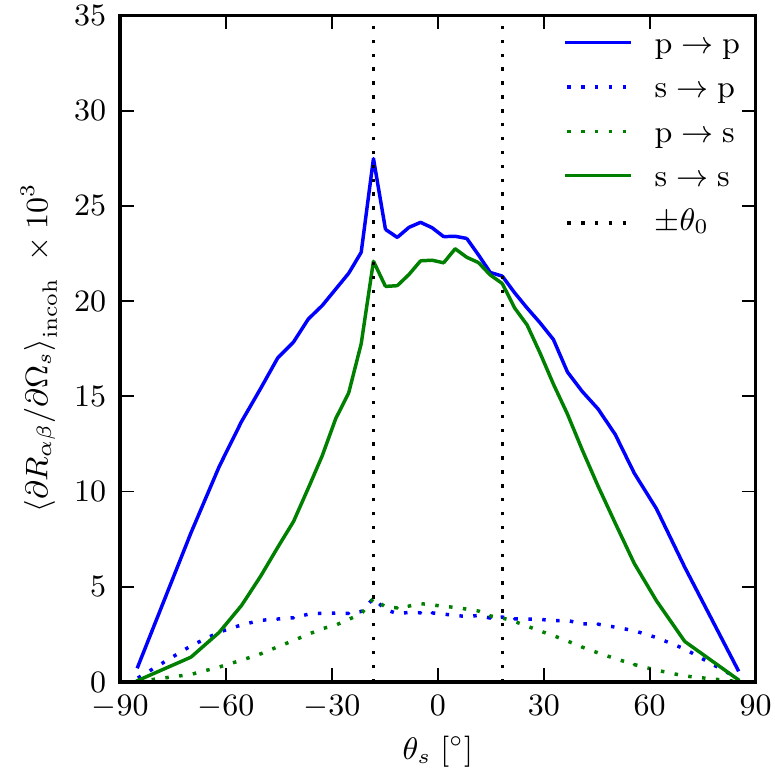}
    \caption{(Color online) Incoherent part of the mean differential reflection coefficient [Eq.~(\ref{eq:drc_incoh})], showing only the in-plane scattering as a function of outgoing lateral wave vector, averaged over 14,200 randomly rough silver surface realizations. The wavelength (in vacuum) of the incident light was $\lambda=\unit{457.9}{\nano\meter}$, and the dielectric function of silver at this wavelength is $\eII=-7.5+0.24\ui$. The surface power spectrum was Gaussian [Eq.~(\ref{eq:gaussian})], with correlation lengths $a_1=a_2=0.25\lambda$ and rms height $\delta=0.025\lambda$. The angle of incidence was $\theta_0=18.24\degree$, the surface covered an area $L \times L$, where $L=25\lambda$, and the surface was discretized on a grid of $319 \times 319$ points. The position of the specular peak (not present in the incoherent part) and the enhanced backscattering peak are indicated by the vertical dashed lines.}
\label{fig:ag_1D_drc}
\end{figure}

\begin{figure}
    \includegraphics[width=\columnwidth]{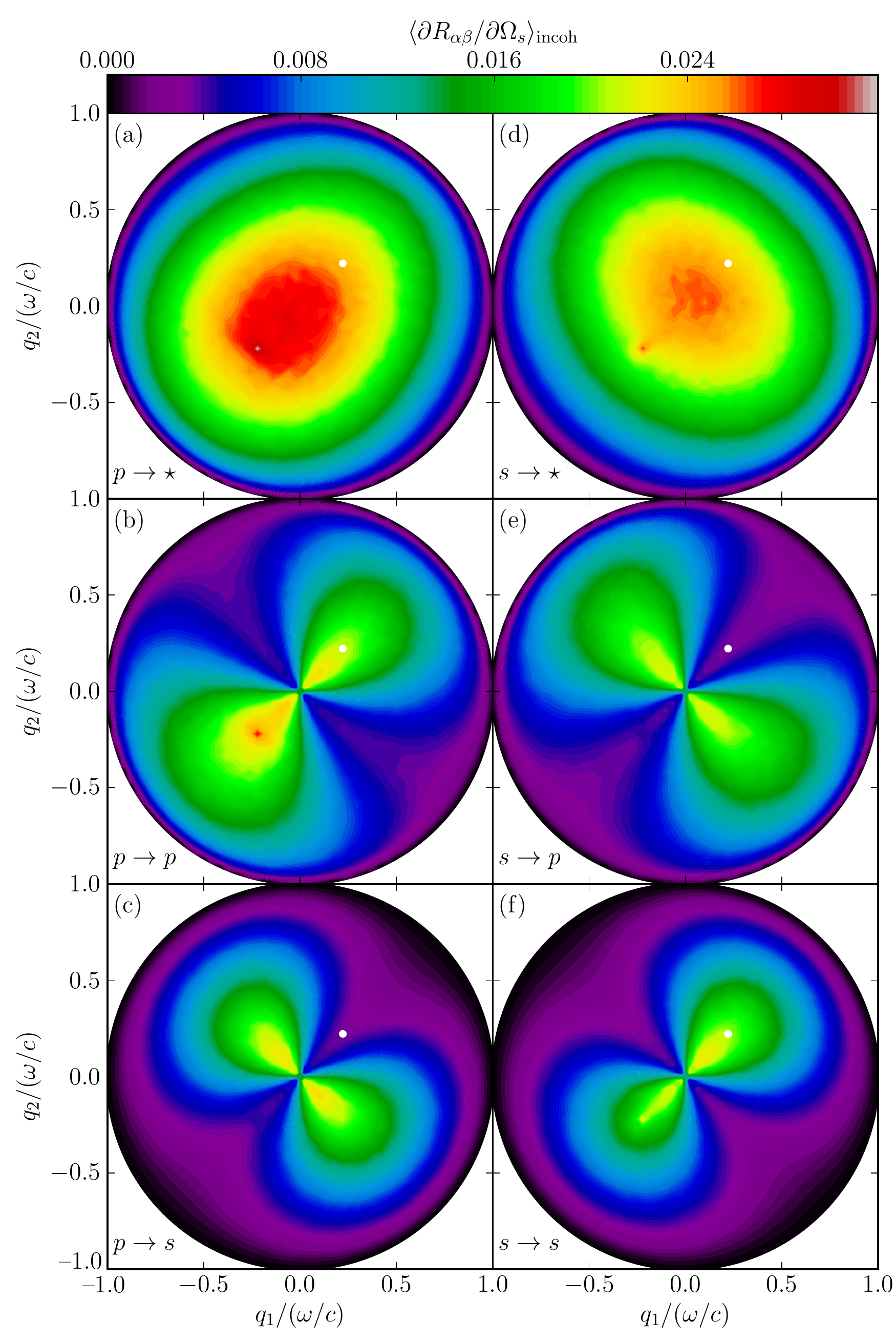}
    \caption{(Color online) Incoherent part of the mean differential reflection coefficient [Eq.~(\ref{eq:drc_incoh})], showing the full angular distribution as a function of outgoing lateral wave vector. All parameters are the same as in Fig.~\ref{fig:ag_1D_drc}. The specular position is indicated by the white dots.}
\label{fig:ag_2D_drc}
\end{figure}

\begin{figure}
    \includegraphics[width=\columnwidth]{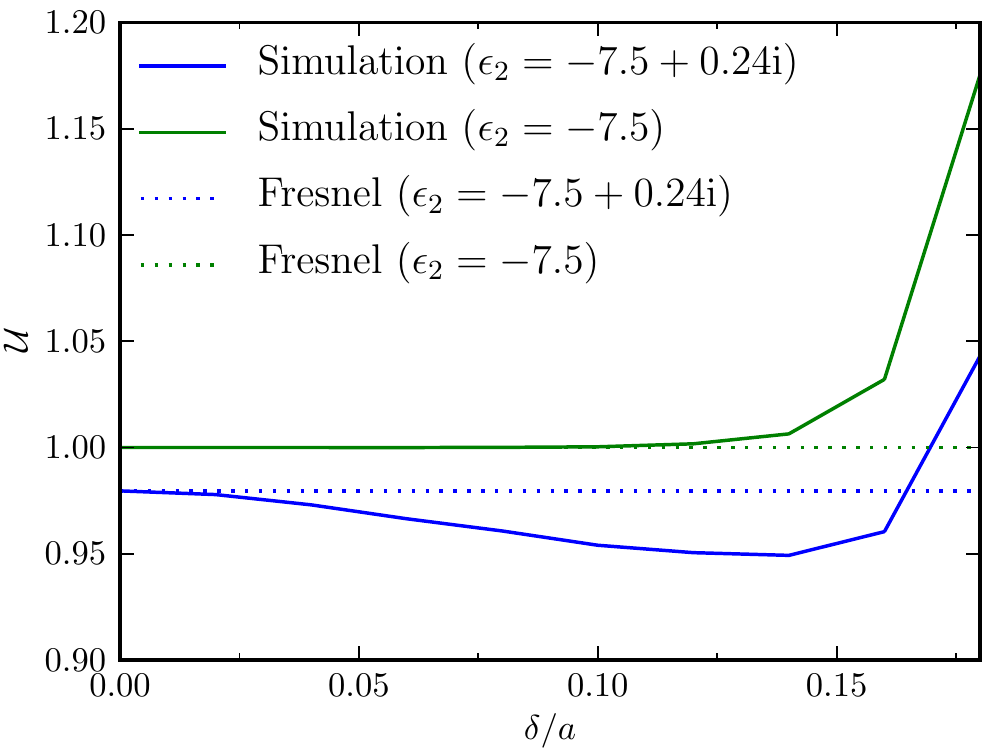}
    \caption{(Color online) Ratio of reflected power to incident power, $\mathcal{U}$, as a function of ratio between rms roughness and correlation length, $\delta/a$. Surface size and resolution were the same as for Fig.~\ref{fig:ag_1D_drc}, and the surface was randomly rough with a Gaussian power spectrum, correlation length was kept constant at $a=a_1=a_2=0.25\lambda$, while the rms roughness $\delta$ was varied from $0.0$ to $0.045\lambda$. The Fresnel coefficients have been included for comparison.}
\label{fig:reflection}
\end{figure}

\begin{figure}
    \includegraphics[width=\columnwidth]{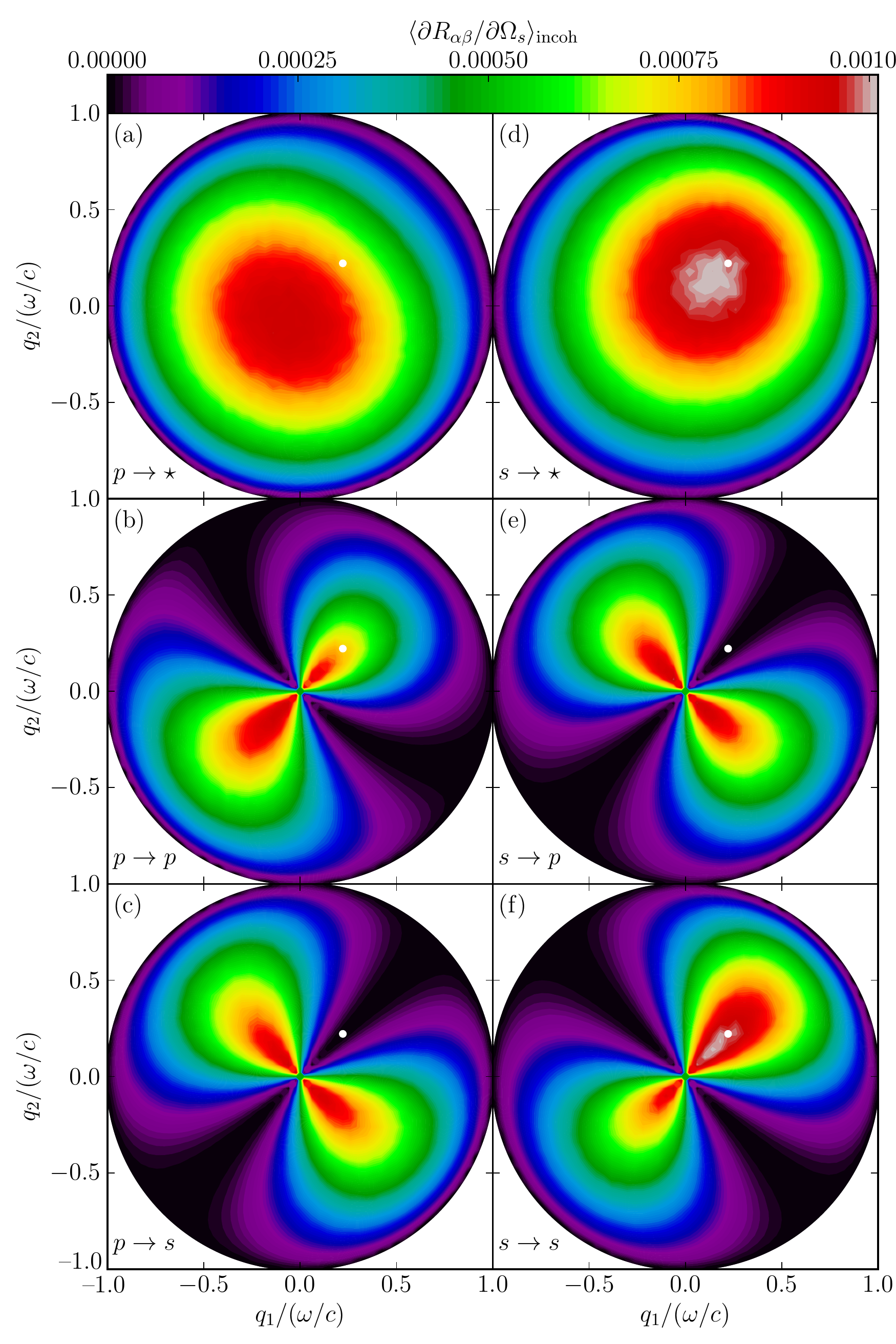}
    \caption{(Color online) The same as in Fig.~\ref{fig:ag_2D_drc}, except that $\eII=2.64$, and the results are averaged over 21,800 randomly rough surfaces.}
\label{fig:di_2D_drc}
\end{figure}

\begin{figure}
    \includegraphics[width=\columnwidth]{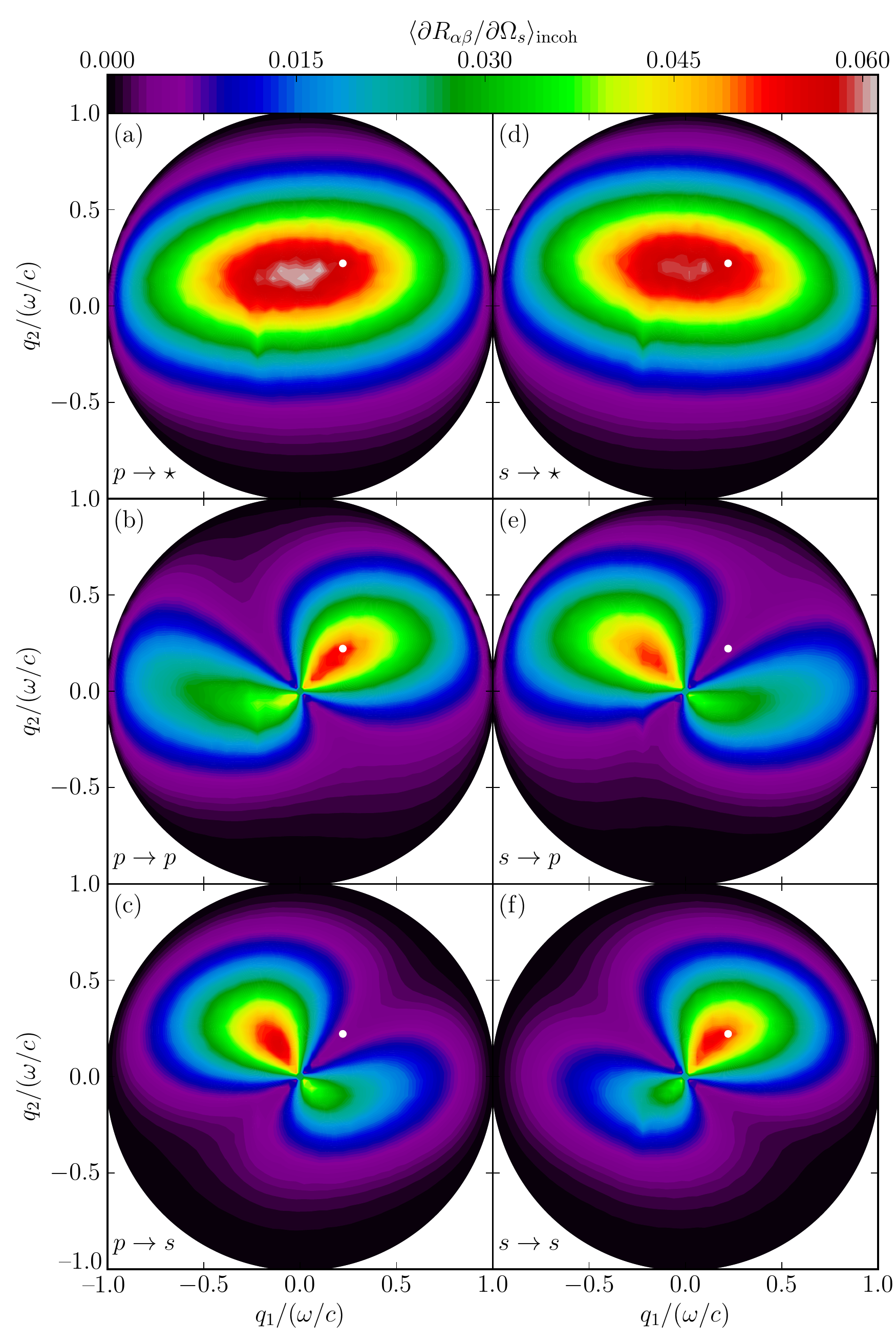}
    \caption{(Color online) The same as in Fig. \ref{fig:ag_2D_drc}, except the correlation length of the Gaussian roughness, which is $a_1=0.25\lambda$ in the $x_1$ direction and $a_2=0.75\lambda$ in the $x_2$ direction, and the results are the average of an ensemble of 6,800 surface realizations.}
\label{fig:ag_anisotropic}
\end{figure}

\begin{figure}
    \includegraphics[width=\columnwidth]{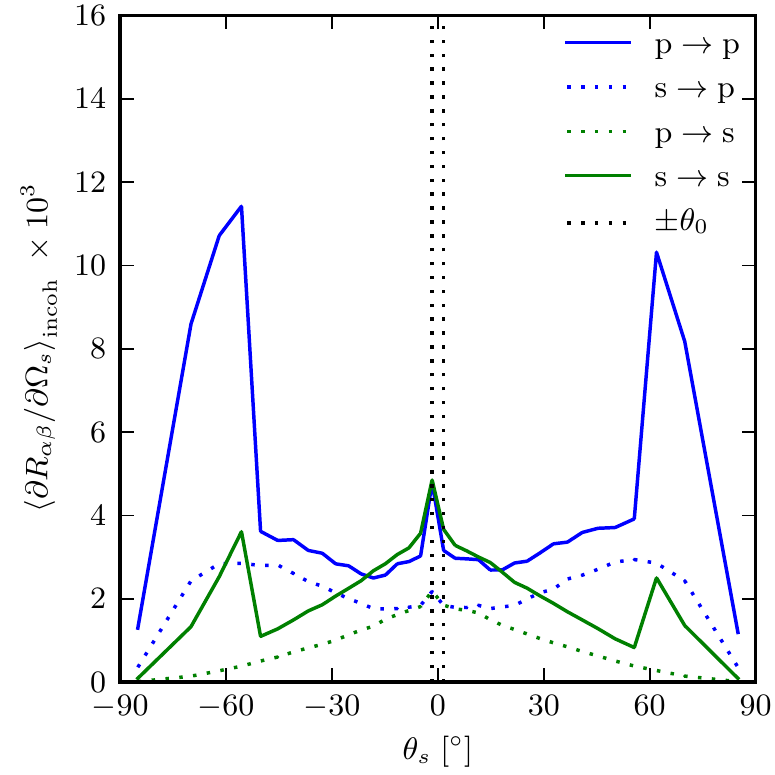}
    \caption{(Color online) Incoherent part of the mean differential reflection coefficient [Eq.~(\ref{eq:drc_incoh})], showing only the in-plane scattering as function of outgoing lateral wave vector, averaged over 7,000 surface realizations with dielectric constant $\eII=-16+1.088\ui$, which corresponds to silver at $\lambda=\unit{632.8}{\nano\meter}$. The surface power spectrum was of the cylindrical type [Eq.~(\ref{eq:cylindrical})], with $k_- = 0.82\omega/c$, $k_+ = 1.97\omega/c$, and rms roughness $\delta=0.025\lambda$. The angles of incidence were $\theta_0=1.6\degree$ and $\phi_0=45\degree$.}
\label{fig:ag_WoD_1D_drc}
\end{figure}

To demonstrate the use of the formalism for solving the reduced Rayleigh equation, the first set of calculations we carried out was for two-dimensional randomly rough silver surfaces. The surface roughness was characterized by an rms height of $\delta=0.025\lambda$ and an isotropic Gaussian power spectrum [Eq.~(\ref{eq:gaussian})] of correlation lengths $a_1=a_2=0.25\lambda$. In Figs.~\ref{fig:ag_1D_drc} and \ref{fig:ag_2D_drc} we present simulation results for the contribution to the mean differential reflection coefficients from light of wavelength (in vacuum) $\lambda=\unit{457.9}{\nano\meter}$ that was scattered incoherently from a rough silver surface of size $25\lambda \times 25\lambda$, discretized into $319\times319$ points. The dielectric function of silver at this wavelength is $\eII=-7.5+0.24\ui$, and the angles of incidence were $\theta_0=18.24\degree$ and $\phi_0=45\degree$.

Figure~\ref{fig:ag_1D_drc} shows the in-plane scattering for this system. The enhanced backscattering peak, a multiple scattering phenomenon, is clearly visible, and is as expected strongest in $\mathrm{p} \to \mathrm{p}$ scattering, since p-polarized light has a stronger coupling to surface plasmon polaritons~\cite{Simonsen-2010}. Figure~\ref{fig:ag_2D_drc} shows the full angular distribution of the mean DRC for the same system. In Figs.~\ref{fig:ag_2D_drc}(a)--(c) and Figs.~\ref{fig:ag_2D_drc}(d)--(e) the incident light was p- and s-polarized, respectively. Figures~\ref{fig:ag_2D_drc}(c) and \ref{fig:ag_2D_drc}(f) show scattering into s-polarization, Figs.~\ref{fig:ag_2D_drc}(b) and \ref{fig:ag_2D_drc}(e) show scattering into p-polarization and in Figs.~\ref{fig:ag_2D_drc}(a) and \ref{fig:ag_2D_drc}(d) the polarization of the scatted light was not recorded. In particular from Fig.~\ref{fig:ag_2D_drc}(b), we observe that the enhancement features seen in Fig.~\ref{fig:ag_1D_drc} at angular position $\theta_s=-\theta_0$, are indeed enhancements in a well-defined direction corresponding to that of retro-reflection, and not some intensity ridge structure about this direction (as has been seen for other scattering systems~\cite{Simonsen2009-1}).
Moreover, the structures of the angular distribution of the intensity of the scattered light depicted in Fig.~\ref{fig:ag_2D_drc} are consistent with what was found by recent studies by using other numerical methods~\cite{Simonsen2009-1,Simonsen2009-9}. 
 The results presented in Figs.~\ref{fig:ag_1D_drc} and \ref{fig:ag_2D_drc} were obtained by averaging the DRC over an ensemble consisting of 14,200 surface realizations.

A test of energy conservation was performed by simulating the scattering of light from a non-absorbing silver surface ($\mathrm{Im}~\eII = 0$) with otherwise the same parameters as those used to obtain the results of Figs.~\ref{fig:ag_1D_drc} and ~\ref{fig:ag_2D_drc}. For this scattering system we found $|\mathcal{U}-1| \leq 0.0003$, i.e., energy is conserved to within 0.03\%, something that testifies to the accuracy of the approach and a satisfactory discretization. 

As a further test, we studied the scattering from a set of (absorbing) silver surfaces with the same parameters used to obtain Figs.~\ref{fig:ag_1D_drc} and \ref{fig:ag_2D_drc}, except that the rms roughness $\delta$ was varied between $0$ and $0.045\lambda$, while the correlation lengths were held constant at $a_1=a_2=0.25\lambda\equiv a$. For the purpose of comparison, we also performed simulations for a similar set of surfaces but assuming no absorption, i.e., we used  $\eII=-7.5$. The results of these tests are presented in Fig.~\ref{fig:reflection}.

The reduced Rayleigh equation is only valid for surfaces of small slopes~\cite{Book:Maradudin-2007}. We have found that at least for the parameters used in obtaining Fig.~\ref{fig:reflection}, our code gives good results for an rms roughness to correlation-length ratio $\delta/a \lesssim 0.12$, as judged by energy conservation. For larger values of $\delta/a$, the results look qualitatively much the same, but the ratio of reflected to incident power starts to become nonphysical (increasing past $1$), as seen in Fig.~\ref{fig:reflection}. It is noted that decreasing the sampling interval $\Delta q$, with $\mathcal{Q}$ unchanged, did not change this conclusion in any significant way, indicating that the observed lack of energy conservation was not caused by poor resolution in discretizing the integral over $\qp$.

The next set of calculations we performed was for a dielectric substrate characterized by $\eII=2.64$. Otherwise, all roughness parameters were the same as for the silver surface used to produce Figs.~\ref{fig:ag_1D_drc} and \ref{fig:ag_2D_drc}. The mean differential reflection coefficient for light scattered incoherently by the rough dielectric surface is presented in Fig.~\ref{fig:di_2D_drc}. By comparing these results to those presented in Fig.~\ref{fig:ag_2D_drc}, we notice that the dielectric reflects less than the silver (the figures show only the incoherent scattering, but the same holds for the coherent part), which is as expected. The ratio of reflected to incident power for these data was $\mathcal{U}=0.0467$ for p-polarized light at an angle of incidence of $\theta_0=18.24\degree$. Moreover, from Fig.~\ref{fig:di_2D_drc} we also notice the absence of the enhanced backscattering peak, which is also to be expected since this phenomenon (for a weakly rough surface) requires the excitation of surface guided modes~\cite{Simonsen-2010}. Note that for a transparent substrate, it is not possible to verify the conservation of energy without also calculating the transmitted field. Therefore, energy conservation has not been tested for the dielectric substrate geometry.

So far, we have exclusively considered surfaces with statistically isotropic roughness. For the results presented in Fig.~\ref{fig:ag_anisotropic}, we simulated the light scattering from a silver surface of the same parameters as those assumed in producing the results of Figs.~\ref{fig:ag_1D_drc} and \ref{fig:ag_2D_drc}, except that now the surface power spectrum was anisotropic, with correlation lengths $a_1=0.25\lambda$ in the $x_1$ direction and $a_2=0.75\lambda$ in the $x_2$ direction and an rms roughness of $\delta=0.025\lambda$. Figure~\ref{fig:ag_anisotropic} shows the incoherent part of the mean DRC averaged over 6,800 surface realizations. In this case, there is more diffuse scattering along the $x_1$ direction than the $x_2$ direction, which is to be expected, since a shorter correlation length means the height of the surface changes more rapidly when moving along the surface in this direction. The interested reader is encouraged to consult Ref.~\cite{Simonsen2010-04} for a more detailed study of light scattering from anisotropic surfaces.

Finally, for the results presented in Fig.~\ref{fig:ag_WoD_1D_drc}, we have simulated the scattering of light from a surface of size $25\lambda \times 25\lambda$, discretized into $319\times319$ points, with $\eII=-16+1.088\ui$, corresponding to silver at a wavelength $\lambda=\unit{632.8}{\nano\meter}$. The surface power spectrum was cylindrical [see Eq.~\eqref{eq:cylindrical}], with $k_- = 0.82\omega/c$, $k_+~=~1.97\omega/c$ and rms roughness $\delta=0.025\lambda$, and the angles of incidence were 
$(\theta_0,\phi_0)=(1.6\degree, 45\degree)$. Figure~\ref{fig:ag_WoD_1D_drc} shows the in-plane, incoherent part of the mean differential reflection coefficient averaged over 7,000 surface realizations. 

From perturbation theory~\cite{Simonsen-2010,Book:Maradudin-2007}, we know that for an incident wave of lateral wave vector $\kp$ to be scattered \emph{via single scattering} into a reflected wave of lateral wave vector $\qp$, we must have $g(\qp-\kp)>0$, where $g(\kp)$ is the surface power spectrum [Eq.~(\ref{eq:powerspectrum})]. Since the power spectrum in this case is zero for \mbox{$|\qp-\kp|<0.82\omega/c$}, we have no contribution from single scattering in the angular interval from $\theta_s=-53.5\degree$ to $\theta_s=56.7\degree$ (for the angles of incidence assumed here). The enhanced backscattering peak, which is due to multiple scattering processes, is clearly visible in Fig.~\ref{fig:ag_WoD_1D_drc} (at $\theta_s=-\theta_0$) partly because it is not masked by a strong single scattering contribution. 


\section{Discussion} 
\label{sec:discussion}

A challenge faced when performing a direct numerical solution of the reduced Rayleigh equation for the scattering of light from two-dimensional rough surfaces is the numerical complexity. In this section, we discuss some of these issues in detail.

\subsection{Memory Requirements} 
\label{sub:memory}

Part of the challenge of a purely numerical solution of the reduced Rayleigh equation by the formalism introduced by this study, is that it requires a relatively large amount of memory. With approximately $\mathcal{N}=(\pi/4) N_q^2$ possible values for $\qp$, the coefficient matrix of the linear equation system will contain approximately $(2\mathcal{N})^2$ elements, where the factor $2$ comes from the two outgoing polarizations. Hence, the memory required to hold the left hand side of the equation system will be approximately $4\mathcal{N}^2\eta$, where $\eta$ is the number of bytes used to store one complex number.

If each element is a single precision complex number, which is what was used to obtain the results presented in this paper, then $\eta=8\;\mathrm{bytes}$, and the matrix will require approximately $2 \pi^2 N_q^4$ bytes of memory for storage. For instance, with the choice $N_x=319$, which was used in all the simulations presented in this paper, and that corresponds to $N_q=160$ [Eq.~\eqref{eq:Nq}], the coefficient matrix will take up approximately 12 GB of memory.

Note that if we instead performed the $\qp$ integration in Eq.~\eqref{eq:RRE} over a square domain of edge ${\mathcal Q}$, the number of elements in the resulting coefficient matrix would be $(2N_q^2)^2=(16/\pi^2)(2{\mathcal N})^2$.  Hence, by restricting the $\qp$ integration present in the reduced Rayleigh equation to a circular domain of radius ${\mathcal Q}/2$, the memory footprint of the simulation is approximately $\pi^2/16\approx0.62$ of what it would have been if a square integration domain of edge ${\mathcal Q}$ was used. For this reason, a circular integration domain has been used in obtaining the results presented in this paper. However, we have checked and found that using a square $\pvec{q}$ integration domain of a similar size will not affect the results in any noticeable way. Indeed, if this was not the case, it would be a sign that $\mathcal{Q}$ was too small.

When determining the system size, we can freely choose the length of the edge of the square surface, $L$, and the number of sampling points along each direction, $N_x$. These parameters will then fix the resolution of the surface, $\Delta x$, the resolution in wave vector space, $\Delta q$, the number of resolved wave vectors, $N_q$, and the cutoff in the $\qp$ integral, $\mathcal{Q}$ [see Eqs.~\eqref{eq:deltax}, \eqref{eq:deltaq}, \eqref{eq:Nq}, and \eqref{eq:Q}]. The combination of $\Delta q$ and $\mathcal{Q}$ then determines the number of resolved wave vectors that actually fall inside the propagating region, $|\qp| < \omega / c$, which is identical to the number of data points used to produce, e.g., Fig.~\ref{fig:ag_2D_drc}.

As we are not free to choose all the parameters of the system, it is clear that some kind of compromise is necessary.  The number of sampling points of the surface along each direction, $N_x$, and how it determines $N_q$ via Eq.~\eqref{eq:Nq}, determines the amount of memory needed to hold the coefficient matrix, as well as the time required to solve the corresponding linear set of equations. Hence, the parameter $N_x$ is likely limited by practical considerations, typically by available computer hardware or time. For a given value of $N_x$, it is possible to choose the edge of the square surface, $L$, to get good surface resolution, at the cost of poor resolution in wave vector space, or vice versa. Note also that changing $L$ will change $\mathcal{Q}$ via $\Delta q$ [see Eqs.~\eqref{eq:deltaq} and \eqref{eq:Q}]. If $\mathcal{Q}$ is not large enough to include evanescent surface modes, like surface plasmon polaritons, multiple scattering will not be correctly included in the simulations, and the results can therefore not be trusted. The optimal compromise between values of $N_x$ and $L$ depends on the system under study.


\subsection{Calculation Time} 
\label{sub:time}

The simulations presented in this paper were performed on shared-memory machines with 24 GB of memory and two six-core 2.4 GHz AMD Opteron processors, running version 2.6.18 of the Linux operating system. The code was parallelized using the MPI library, and the setup of the linear set of equations ran on all 12 cores in the timing examples given. The linear equation solver used was a parallel, dense solver based on LU-decomposition~\cite{Book:NR-1992} (PCGESV from ScaLAPACK), which runs efficiently on all 12 cores. Setting up the equation system scaled almost perfectly to a large number of cores, while the solver scaled less well, due to the need for communication.  Numerically solving the reduced Rayleigh equation for the scattering amplitudes associated with one realization of a rough surface, discretized onto a grid of $319\times319$ points, took approximately 17 minutes on the architecture described above, and required about 12 GB of memory. Out of this time, approximately 100 seconds was spent setting up the equation system, 950 seconds was spent solving it by LU decomposition, and typically around 1 second was spent on other tasks, including writing data to disk. Table~\ref{tab:time} shows timing and memory details of the calculations, including other system sizes.

Based on the discussion in Sec.~\ref{sub:memory}, we note that the use of a circular $\qp$ integration domain also significantly reduces the time required to solve the resulting linear system of equations. When using a dense solver, the time to solve the systems scales as the cube of the number of unknowns. Thus we expect the CPU time to solve the matrix system for a circular integration domain of radius $\mathcal{Q}/2$ to be about half ($\pi^3/2^6$) the time to solve the corresponding system using a square domain of edge $\mathcal{Q}$.

The ratio of the time spent solving one equation system to the total simulation time per surface realization increases with increasing system size, as the time to set up the equation system is $\mathcal{O}(N_x^4)$, while the time to solve the linear system by LU decomposition scales as $\mathcal{O}(N_x^6)$. It is clear from Table~\ref{tab:time} that for any surface of useful size the runtime is completely dominated by the time spent in solving the linear set of equations. 

Since the time solving the equation system dominates the overall simulation time, we investigated if one could improve the performance of the simulations by using an iterative solver instead of a direct solver based on LU decomposition. For example, Simonsen et al.~\cite{PhysRevA.81.013806} recently reported good performance using BiCGStab~\cite{vorst:631} on a dense matrix system of a similar size. In our preliminary investigations into using iterative solvers, we found that convergence with BiCGStab was slow and unreliable for our linear equation systems. However, it should be stressed that we did not use a preconditioning scheme, which could potentially yield significantly improved convergence.

From Eq.~(\ref{eq:RRE}) it follows that changing the angles of incidence and/or the polarization of the incident light changes {\em only} the right hand side of the equation system to be solved. Hence, an advantage of using LU decomposition (over iterative solvers) is that the additional time required to solve the system for several right hand sides is negligible, since the overall majority of time is spent factorizing the matrix. Conversely, the time spent using an iterative solver (like BiCGStab) will scale linearly with the number of right hand sides. For these reasons, we have chosen to use an LU-based solver.

\begin{table}
  \caption{\label{tab:time}
    Walltime spent to solve the RRE for various values of $N_x$ on a shared-memory machine with two six-core 2.4 GHz AMD Opteron processors. Included are total time ($t_{\mathrm{tot}}$), time to setup the coefficient matrix of the equation system ($t_{\mathrm{LHS}}$) and the time to solve the equation system ($t_{\mathrm{solve}}$). Also included is the memory required to store the coefficient matrix of the linear equation system for each run ($M_{\mathrm{LHS}}$). The time to set up the right hand side of the linear equation system is negligible compared to the left hand side, and have therefore not been included here.}
\begin{ruledtabular}
    \begin{tabular}{rdddd}
              \multicolumn{1}{c}{$N_x$}
            & \multicolumn{1}{c}{$t_{\mathrm{LHS}}(\second)$}
            & \multicolumn{1}{c}{$t_{\mathrm{LU}}(\second)$}
            & \multicolumn{1}{c}{$t_{\mathrm{tot}}(\second)$}
            & \multicolumn{1}{r}{$M_{\mathrm{LHS}}(\mathrm{GB})$} \\
        \hline
        199 & 10 & 58 & 69 & 1.8 \\
        239 & 28 & 171 & 200 & 3.8 \\
        279 & 56 & 429 & 486 & 7.0 \\
        319 & 97 & 946 & 1,045 & 12.0 \\
        369 & 154 & 1,916 & 2,074 & 19.2 \\
        399 & 266 & 3,625 & 3,895 & 29.4 \\
    \end{tabular}
\end{ruledtabular}
\end{table}


\subsection{GPU implementation} 
\label{sub:GPU implementation}

Currently, performing simulations like those presented in this paper on a single desktop computer is prohibitively time consuming due to inadequate floating point performance. However, the increasing availability of powerful graphics processing units~(GPUs) has the potential to provide computing power comparable to that of a powerful parallel machine, but at a fraction of the cost. As the most time-consuming step in our simulations is the LU decomposition of the system matrix (see Table~\ref{tab:time}), this is where efforts should be made to optimize the code. With this in mind, the simulation code was adapted to (optionally) employ version 1.0 of the MAGMA library~\cite{agullo2011lu} for GPU-based LU decomposition. Performance was compared between a regular supercomputing service and a GPGPU (General Purpose GPU) testbed. On the regular service, the code was running on a single compute node containing two AMD 2.3 GHz 16-core processors and 32 GB of main memory. On the GPGPU testbed, the hardware consisted of a single Nvidia Fermi C2050 processor with 3 GB of dedicated memory and 32 GB of main system memory. For these two computer systems, the initial performance tests indicated that the LU decomposition took comparable time on the two architectures for a system of size $N_q = 100$ (the difference was less than 10\%). The time using the GPGPU testbed included the transfer of the system matrix to and from the Fermi card and the decomposition of the matrix. Even though these results are preliminary, it demonstrates that there is a possibility of performing simulations like those reported in this study without having to resort to costly supercomputing resources. Instead, even with limited financial means, they may be performed on single desktop computers with a state-of-the-art GPU.


\section{Conclusion} 
\label{sec:Conclusion}
We have introduced a formalism for performing non-perturbative, purely numerical, solutions of the reduced Rayleigh equation for the reflection of light from two-dimensional penetrable rough surfaces, characterized by a complex dielectric function $\varepsilon(\omega)$.

As an example, we have used this formalism to carry out simulations of the scattering of p- or s-polarized light from two-dimensional randomly rough dielectric and metallic surfaces characterized by isotropic or anisotropic Gaussian and cylindrical power spectra. From the scattering amplitudes, obtained by solving the reduced Rayleigh equation, we calculate the mean differential reflection coefficients, and we calculate the full angular distribution of the scattered light, with polarization information. For the scattering of light from weakly rough metal surfaces, the mean differential reflection coefficient shows a well-defined peak in the retro-reflection direction (the enhanced backscattering phenomenon). From previous experimental and theoretical work, this is to be expected for such scattering systems. Moreover, the obtained angular distributions of the intensity of the scattered light show the symmetry properties found for strongly rough surfaces in recent studies using other simulation methods.

For the purpose of evaluating the accuracy of our simulation results, we used the conservation of energy for a corresponding non-absorbing scattering system. This is a required, but not sufficient, condition for the correctness of the numerical simulations. By this method, we found that within the validity of the reduced Rayleigh equation our code produces reliable results, at least for the parameters assumed in this study. In particular, for a rough non-absorbing metal surface of the parameters used in this study, energy was conserved to within $0.03\%$, or better. This testifies to the accuracy of the approach and a satisfactory discretization. Moreover, we also performed simulations of the scattered intensity for systems where the rms roughness of the surface was systematically increased from zero with the other parameters kept unchanged. It was found that energy conservation was well satisfied (for the parameters assumed) when the ratio of rms roughness~($\delta$) to correlation length~($a$), satisfied $\delta/a \lesssim 0.12$.

We believe that the results of this paper provide an important addition to the collection of available methods for the numerical simulation of the scattering of light from rough surfaces. The developed approach can be applied to a wide range of scattering systems, including clean and multilayered scattering systems, that are relevant for numerous applications.


\acknowledgments
We would like to acknowledge the help of Dr. Chris Johnson at the EPCC, University of Edinburgh, for help in parallelizing the code. We are also indebted to Dr. A. A. Maradudin for discussions on the topic of this paper. The work of T.N. and P.A.L. was partially carried out under the HPC-EUROPA2 project (project number: 228398) with support of the European Commission -- Capacities Area -- Research Infrastructures. The work was also supported by NTNU by the allocation of computer time.

\bibliography{references}

\end{document}